# DEVELOPMENT OF AN ANTI-COLLISION MODEL FOR VEHICLES

Adamu Murtala Zungeru

School of Electrical and Electronic Engineering, University of Nottingham, Jalan Broga, 43500 Semenyih, Selangor Darul Ehsan, Malaysia
keyx1mzd@nottingham.edu.my

## ABSTRACT

*The Anti–Collision device is a detection device meant to be incorporated into cars for the purpose of safety. As opposed to the anti–collision devices present in the market today, this system is not designed to control the vehicle. Instead, it serves as an alert in the face of imminent collision. The device is intended to find a way to implement a minimum spacing for cars in traffic in an affordable way. It would also achieve safety for the passengers of a moving car. The device is made up of an infrared transmitter and receiver. Also incorporated into it is an audio visual alarm to work in with the receiver and effectively alert the driver and/or the passengers. To achieve this design, 555 timers coupled both as astable and monostable circuits were used along with a 38 KHz Square – Pulse generator. The device works by sending out streams of infrared radiation and when these rays are seen by the other equipped vehicle, both are meant to take the necessary precaution to avert a collision. The device would still sound an alarm even though it is not receiving infrared beams from the oncoming vehicle. This is due to reflection of its own infrared beams. At the end of the design and testing process, overall system was implemented with a constructed work, tested working and perfectly functional.*

## 1. INTRODUCTION

Safety is a necessary part of man's life. Due to the accident cases reported daily on the major roads in all parts of the developed and developing countries, more attention is needed for research in the designing an efficient car driving aiding system. It is expected that if such a device is designed and incorporated into our cars as a road safety device, it will reduce the incidence of accidents on our roads and various premises, with subsequent reduction in loss of life and property.

However, a major area of concern of an engineer should be safety, as it concerns the use of his/her inventions and the accompanying dangers due to human limitations. When it comes to the use of a motor vehicle, accidents that have occurred over the years tell us that something needs to be done about them from an engineering point of view. According to the 2007 edition of the Small-M report on the road accident statistic in Malaysia [1], a total of 6,035 people were killed in 2000 and the fatality spring up to 6,287 in 2006 from accident cases reported in 250,429 and 341,252 cases of accident for 2000 and 2006 respectively. These road accidents were mainly at Kelang and KL area of Malaysia. The obtained results show that, high rate of accident is reported each year [1].

Suffice to say that the implementation of certain highway safety means such as speed restrictions, among others, has done a lot in reducing the rates of these accidents. The issue here is that policies of safe driving alone would not eradicate this, the engineer has a role to play, after all the main issue is an engineering product (the motor vehicle). Many motorists have had to travel through areas with little light under much fatigue, yet compelled to undertake the journey out of necessity. It is not always irresponsible to do this. A lot of cases reported is as a result of drivers sleeping off while driving, and when he/she eventually woke up, a head-on collision might have taken place. Not many have had the fortune to quickly avert this. It is therefore imperative to consider the advantages of an early warning system where the driver is alerted of a possible collision with some considerable amount of time before it occurs.

The idea of incorporating radar systems into vehicles to improve road traffic safety dates back to the 1970s. Such systems are now reaching the market as recent advances in technology have allowed the signal processing requirements and the high angular resolution requirements from physically small antennas to be realized. Automotive radar systems have the potential for a number of different applications including adaptive cruise control (ACC) and anti-collision devices [2]. The problem with this brand of cars is that they are expensive. This becomes an even bigger challenge when you consider a developing country like Malaysia. A real example of such cars in metropolitan areas is shown below in Fig. 1.

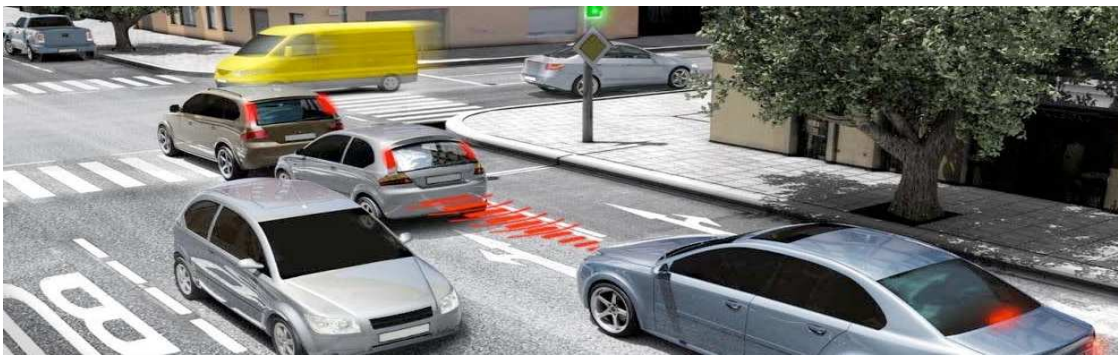

(a)

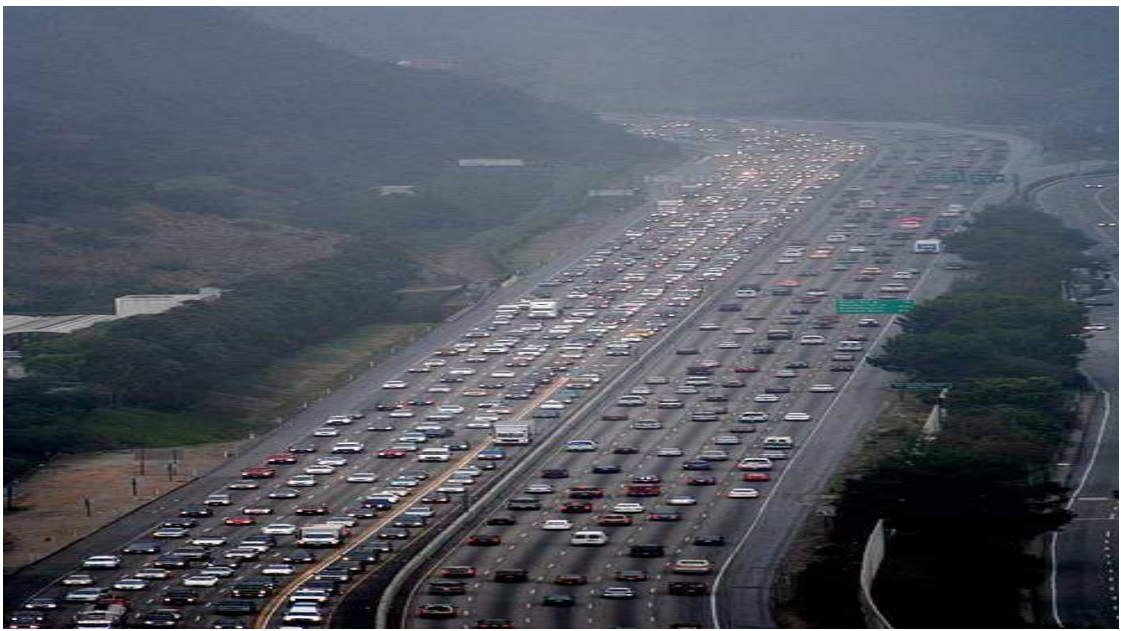

(b)

Fig. 1: High traffic loads in metropolitan areas.

The Infrared Anti-Collision Device are expected to be made of relatively inexpensive components for easy purchase and incorporation. This research aims at the design of a prototype showing how this could function. The main objective is to find a way to implement a minimum spacing for cars in traffic in an affordable way, alongside to achieve safety for passengers of a moving car. The anti-collision device, when wired into the circuitry of a vehicle would help in the reduction of road mishaps. Though not every kind of collision can be helped by this, and it must be stated here that no allusion is being made that technology is the best line of action to take. It should be further noted that some already existing laws made use of technologies like the street lights and traffic lights. This would be a supplementation and not a replacement.

For clarity and neatness of presentation, the article is presented in 5 sections. Section 1 gives a general introduction of anti-collision devices. Theoretical background and review of system components used for this system design is presented in Section 2. Section Three outlined the design and implementation procedures of the anti-collision model. In Section 4, we present the experimental results and follow by analysis and discussion of the results gotten. Conclusions and recommendations are presented in Section 5. Finally, the references are presented at the end of the paper.

## 2. REVIEW AND THEORETICAL BACKGROUND

The idea of using infrared signals to establish routes in communication networks between receivers and transmitters for the purpose of convenience, safety and guarantee of service is not new, but the application, cost, design method and reliability of the system varies. Besides, much were treated in papers by Zungeru et al. [3, 4]. In their papers, the use of infrared rays was studied and utilized to count the number of passengers in a car and also remotely control home appliances via short message services. Generally, the anti–collision device prototype designed here is a detection device, sensitive to solid objects in its pathway. To achieve this sensitivity, some of the basic components used for its realization are discussed in Section 2.1.

### 2.1. Theoretical Background

A basic makeup of the device is an infrared transmitter/receiver and an audio/visual alarm using two different colors of LED and a speaker. Each was duplicated for better demonstration. The system was designed around three basic modules: (1) High Frequency 30KHz square-wave oscillator, (2) Monostable multivibrators and (3) A D.C. Battery power supply. These different modules are discussed below in detail in the subsequent sub-sections.

### 2.2. High Frequency Square – Wave Oscillator

The infrared sensor type used for this system demands that a modulated infrared source be used for the sensing circuit to effectively respond as required. Infrared sensors are specified for three different modes of operation, and these are explained next.

#### 2.2.1. Mode 1

The amplitude and carrier frequencies are constant here. The sensor output goes momentarily low as illustrated in Fig. 2, and this happens just once during the initial period. For the output to again go low the continuous carrier wave will have to be removed, else, the detector goes blind leaving its output on a high.

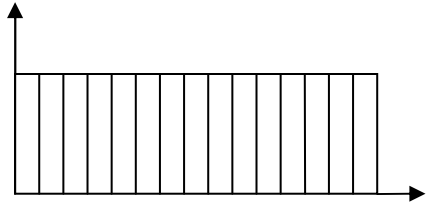

Fig. 2: Oscillator Output for 40 KHz Carrier.

### 2.2.2. Mode 2

The sensor output goes briefly lower each time a 40 KHz pulse is received. The duration of the output pulse is shorter than that of the 40 KHz signal. This is shown in Fig. 3 below.

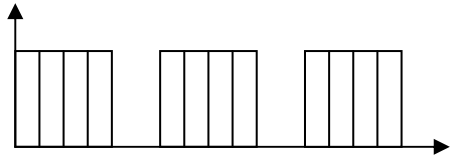

Fig. 3: 40 KHz Carriers Pulsed at 7Hz.

### 2.2.3. Mode 3

Here the output from the detector produces 833 Hz pulses at a frequency rate of 7 Hz. The 833 Hz pulses represent the data that would be sent to the receiver.

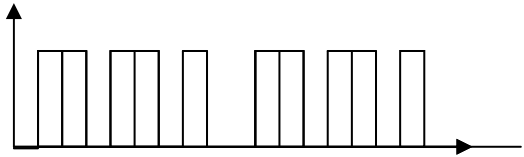

Fig. 4: 40 KHz Carrier modulated at 833Hz and Pulsed at 7Hz.

### 2.3. Monostable Multivibrator

It is also called a "single – shot", "single – swing" or "one – shot" multivibrator. It has one capacitor with which it stores charge [5]. Below is a diagram showing the constituents of a Monostable Multivibrator.

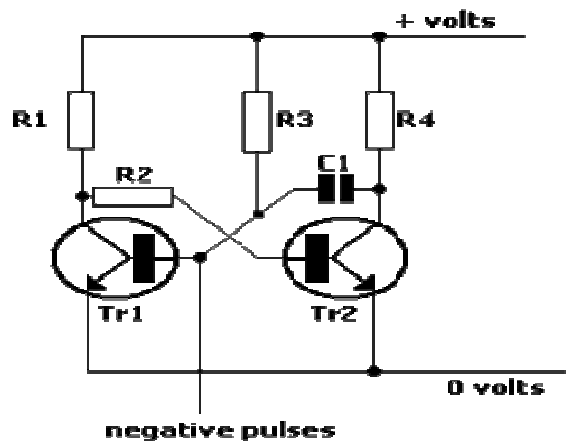

Fig. 5: A Monostable Multivibrator

The monostable is also called a "One – Shot" [6]. One – shots are available in integrated circuit form even though they can be built from the general purpose op–amps [7]. It is triggered by a rising or falling edge at the appropriate input. The only requirement on the triggering signal is that it has some minimum width. The pulse width can be longer or shorter than the output pulse. Most monostables can be retriggered, beginning a new timing cycle, if the input does the triggering in the duration of the output pulse. These are termed Retriggerable Monostables. Some monostables are non–retriggerable; they ignore input transitions during the output pulse. An example is the 555 Timer device used in this work.

The 555 chip is configured in the monostable timer mode when wired as indicated in Fig. 6 below:

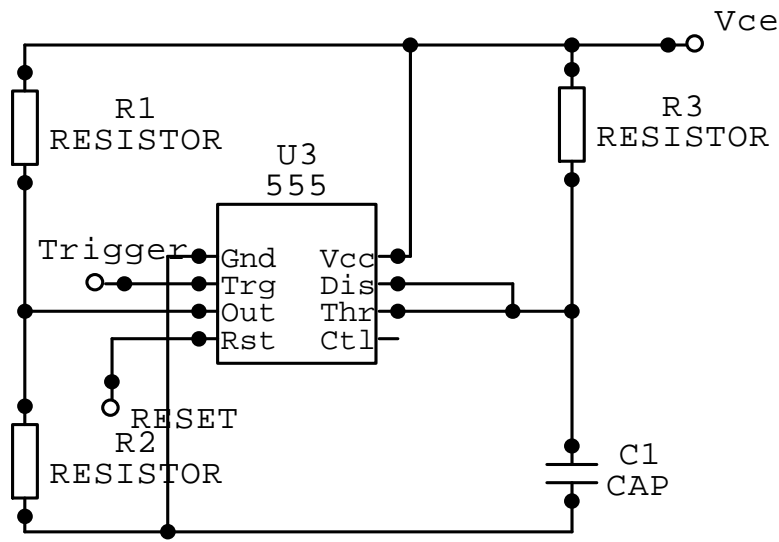

Fig. 6: A 555Timer in a Monostable

The external capacitor, $C_T$, is initially held discharged by a transistor inside the timer. Upon application of a negative trigger pulse of less than 1/3 $V_{CC}$ to Pin 2, a flip – flop is set. This both releases the short – circuit across the capacitor $C_T$ and drives the output high. The voltage across the capacitor then increases exponentially for a period of time; $t = 1.1\ R_T C_T$, at the end of which the voltage becomes equal to 2/3 $V_{CC}$. An external capacitor normally resets the oscillator, which in its further effect discharges the capacitor and helps in taking the output to low state. Since the charge and the threshold level of the capacitor are both directly proportional to supply voltages, the timing interval is independent of supply. During the timing cycle when the output is high, further application of a trigger will not affect the circuit so long as the trigger input is returned high at least 10 microseconds before the end of the timing interval. Though, the circuit can be reset during this time by the application of a negative pulse to the reset pin, which is normally the use of Pin 4 of the timer. and this output will definitely remain in its low state until an external trigger pulse is applied. The timing delay generated using a 555 monostable is:

$T_d = 1.1\ R_T C_T$ (1)

Where $R_T$ is the resistance between the supply and pins 6 and 7; and $C_T$ is the capacitance between Pins 6 and 7 and ground.

The IR sensor output are OR'ed and used at the enable input of a 555 oscillator (de - asserting reset), generating a 250Hz audio tone. The output from the monostable independently controls visual LED indicators that show the origin of distance violation [7].

## 2.4. Astable Multivibrator

It is a free – running relaxation oscillator. It has no stable state but only two quasi – stable (half – stable) between which it keeps oscillating of its own accord without external excitation. In this circuit, neither of the transistors reaches a stable state, their frequency of change between the ON/OFF states is determined by the RC time constant in the circuit. Two outputs ($180^0$ out of phase) are available [5, 7].

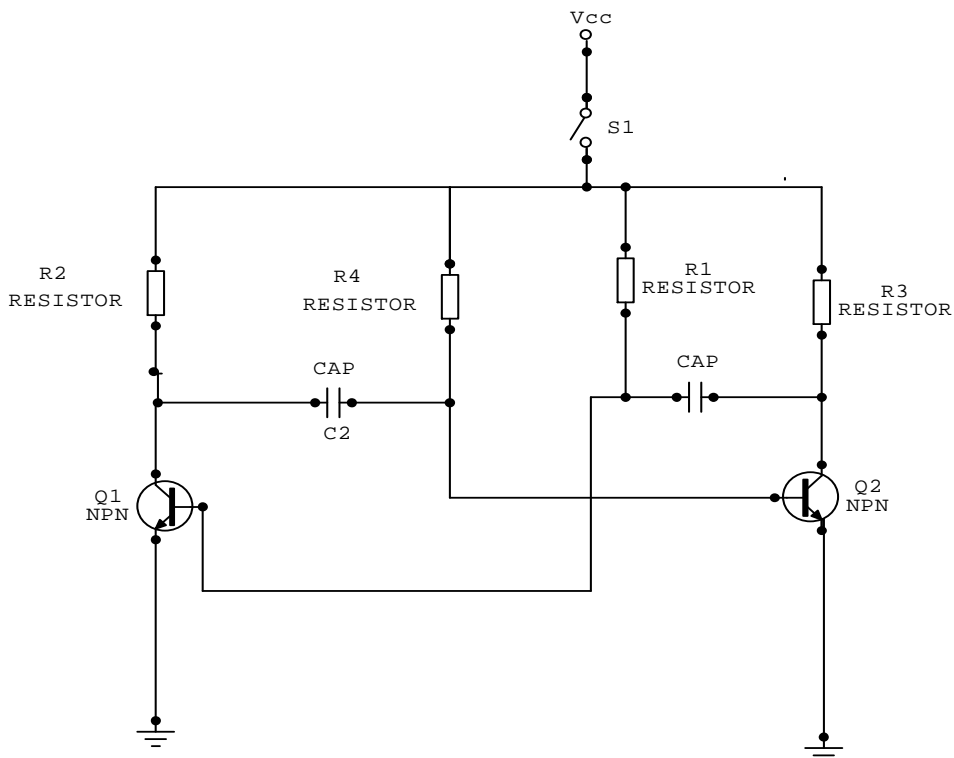

Fig. 7: Astable multivibrator

The 555 Astable Oscillator is a form of relaxation oscillator in which a current source changes a capacitor, then discharging it rapidly when the voltage reaches some threshold, beginning the cycle anew. The operation of the 555 Astable Oscillator is explained as: The output pin (3), goes high when the 555 receives a trigger input, on Pin 2, and it stays there until the threshold (Pin 6) is driven, at which time the output goes low and the discharge transistor (collector on Pin 7) is turned on. The trigger input is activated by an input level below 1/3 $V_{CC}$, and the threshold is activated by that above 2/3 $V_{CC}$. When power is applied to a 555 Astable, the capacitor is discharged, so the 555 is triggered causing the output to go high, the discharge transistor to turn off, and the capacitor to begin charging towards 2/3 $V_{CC}$ through $R_A + R_B$. When it reaches this value, the threshold input is triggered, causing the constant to go low and the discharge transistor to turn ON, discharging C towards the ground through $R_B$. The operation is now cyclic, with the voltage at C going between 1/3 $V_{CC}$ and 2/3 $V_{CC}$, with period T = 0.693 ($R_A + 2R_B$) C.

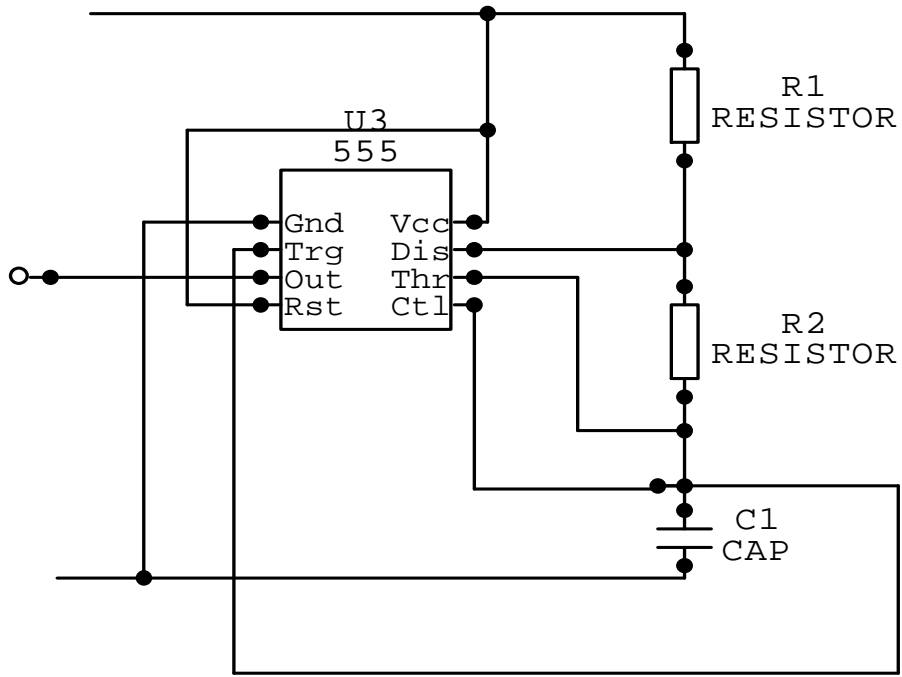

Fig. 8: A 555 Timer Astable Multivibrator

**2.5. Other Types of Detectors**

These are an example of other types of detection objects such as: (1) Radar Detectors, one kind of detector is a radar type detector, and Radar is an acronym for Radio Detection and Ranging. Firstly, its use as a speed gun is considered. A basic speed gun is a radio transmitter and receiver combined into one unit. A radio transmitter is a device that oscillates an electric current so the voltage goes up and down at a certain frequency [9]; (2) LIDAR Guns [10]; Infrared Radiation [11].

## 3. SYSTEM DESIGN AND IMPLEMENTATION

A basic makeup of the device is an infrared transmitter/receiver and an audio/visual alarm using two different colors of LED and a speaker. Each was duplicated for better demonstration. The project was designed around three basic modules: (1) High Frequency 30Khz square-wave oscillator, (2) 555 Monostable and (3) D.C. Battery power supply

**3.1. High Frequency Square-Wave Oscillator**

Since most commercially available three – terminal infrared sensors are most sensitive in the 38Khz region, particularly the TSOP1738 used, a stable 38Khz source is mandatory. A critical examination of the various configurations shows that only mode 2 or mode 3 type of operation is applicable in the implementation of the desired system. The mode 2 type of operation is employed here, as this gives the basic implementation of the infrared (IR) link between vehicles.

A 38 KHz square – wave oscillator is modulated by a low frequency 7Hz source to generate a pulsed output at the sensor's end. This low modulating frequency allows for easy visual observation as indicator Light Emitting Diodes (LEDs) blink in response to the received IR radiation. The charge time (output high) is given by:

$$t_1 = 0.693 \ (R_A + R_B) \ C \qquad (2)$$

And the discharge time (output low) is given by:

$$t_2 = 0.693 (R_B) C \tag{3}$$

Thus the period is:

$$T = t_1 + t_2 \tag{4}$$

$$\text{That is; } T = 0.693 (R_A + 2R_B) C \tag{5}$$

The frequency of oscillation is; $f = 1/T = 1/[0.693 (R_A + 2R_B) C]$

$$\text{Therefore } f = 1.44/ [(R_A + 2R_B) C] \tag{6}$$

The duty cycle is: $D = R_B/ (R_A + 2R_B)$

Since the frequency of operation of the three – terminal IR sensor used lies between 32 – 40 KHz, the oscillator frequency must therefore lie in this range. Most sensors are optimally responsive at 38 KHz as such this frequency is more often used in IR systems. Since modulation is also desired, a means of generating a modulated signal is required. The 555 device has an extra input pin (Pin 5 – modulation input) which allows an input waveform to directly modulate the free running 38 KHz astable frequency. This frequency modulation produces a 38 KHz IR carrier modulated at 7 Hz. This effectively puts the transmitter–sensor interface in mode 2.

Using $f = 1.44/ [(R_A + 2R_B) C]$ and an output frequency of 38 KHz;

$38000 = 1.44/ [(R_A + 2R_B) C]$;

A standard capacitance value of 0.0015μF was chosen and $R_A$ = 2.2 KHz. Substituting both values into the equation above yields 11.53 KΩ.

This was achieved using a variable potentiometer of 20 KΩ.

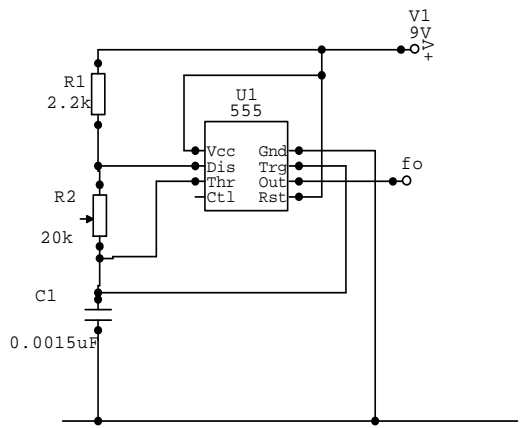 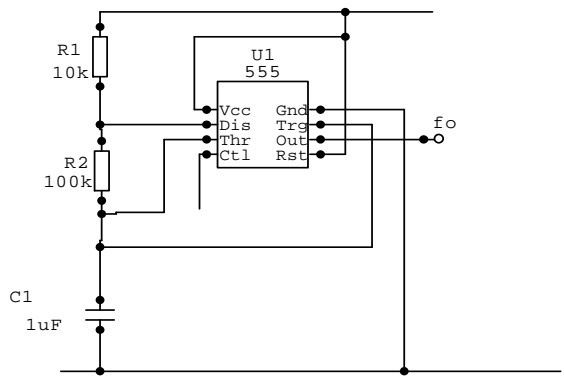

Fig. 9: Oscillator A             Fig. 10: Oscillator B

The modulation source is coupled as shown in Fig. 10. Both oscillators are shown wired together as below (Fig. 11):

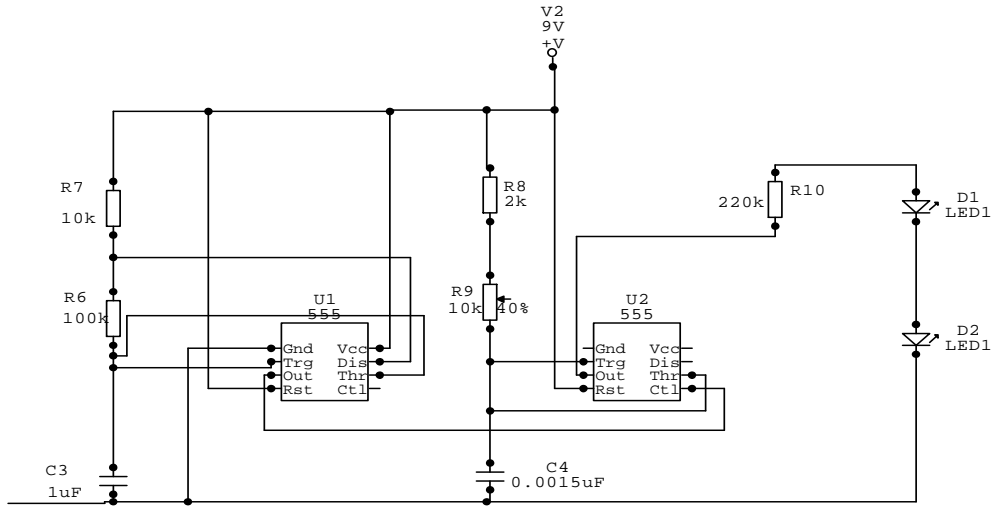

Fig. 11: Coupling of Both Oscillators

The modulated 38 KHz carrier signal drives two series connected IR LED through a 200Ω resistance in series with the diodes.

$R_S$ is deduced from the equation; $R_S = (V_{CC} - V_{LED})/ I_{LED}$

Since the diodes are connected in series, the same current, $I_{LED}$, flows through them.

$V_{CC}$ = 9V, and $V_{LED}$ = 1.7V (measured), $I_{LED}$ has a maximum 0.5A for the diodes used.

Using these values, $R_{S(min)}$ = [9-2(1.7)]/0.5 = (9 – 3.4)/0.5

Reducing $I_{LED}$ ensures diode longevity. A 220Ω resistance was used to yield a low $I_{LED}$, and a low power IR carrier since the distance of travel of the generated carrier is a direct function of the diode forward current.

**3.2. The Receiver Circuit**

The receiver circuit comprises two monostables, as explained earlier on, and a low frequency audio oscillator designed around the 555 timer IC. Two IR sensors connected to two different monostables produce a common output logic that enables/disables the audio frequency oscillator. Each monostable responds to the IR beam sent in its direction by generating a pulsed 1 second square-wave output. This output forms the input to the transistor equivalent.

**3.3. Analysis**

The two monostables provide a one second output that Gates ON or OFF the audio oscillator. Besides the gating action, each monostable also drives a visual indicator that shows the origin of the minimum inter–car distance violation. When the minimum spacing distance is breached from either direction, front or rear, the appropriate sensor picks up the IR beam transmitted from the offending car, turns the appropriate visual indicator ON to reflect the direction of violation. At the same time, it enables the audio frequency oscillator that provides enough power to drive a loudspeaker. Since this is merely a detection device, the occupants of both cars will be notified of the need to adjust to a safe distance apart.

Two SC1815GR NPN transistors are wired in parallel with their collectors common and pulled to +9V via a 10KΩ resistance. This 10KΩ resistance is also the $R_B$ for a common–emitter NPN transistor whose collector load is a 10KΩ resistance as depicted below in Fig. 12. With transistors $Q_1$ and $Q_2$ OFF, $R_3$ sources current into the base emitter direction of $Q_3$. This current is high enough to saturate the transistor, putting its collector at a potential that is just a few tenths of a volt above the emitter potential, 0V.

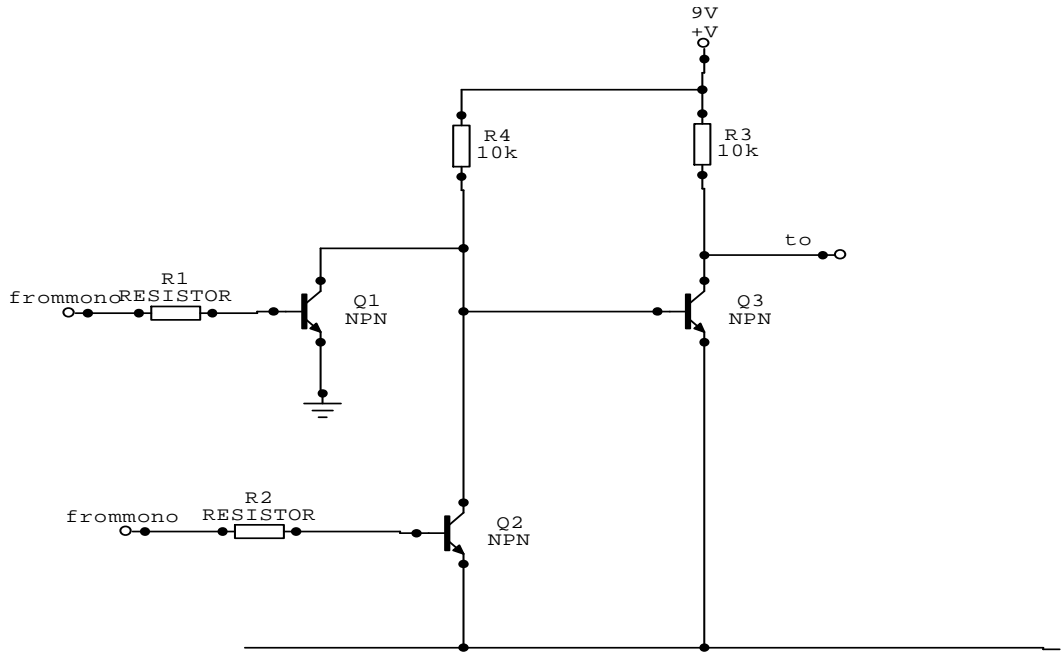

Fig. 12: Oscillator connection with Receiver

Since the $Q_3$'s collector is connected to RESET on the audio oscillator, and with $Q_3$ saturated, RESET is active. Correspondingly, the oscillator is disabled and no tone is generated. If either $Q_1$ or $Q_2$ is ON (driven by the associated monostable), the base of $Q_3$ is shorted to ground, forcing it OFF. The $Q_3$'s collector voltage rises to approximately $V_{CC}$ de-asserting RESET. The audio oscillator is now enabled and it generates a frequency,

$$F = 1.44/[(R_A + 2R_B) C] \text{ Hz} \tag{7}$$

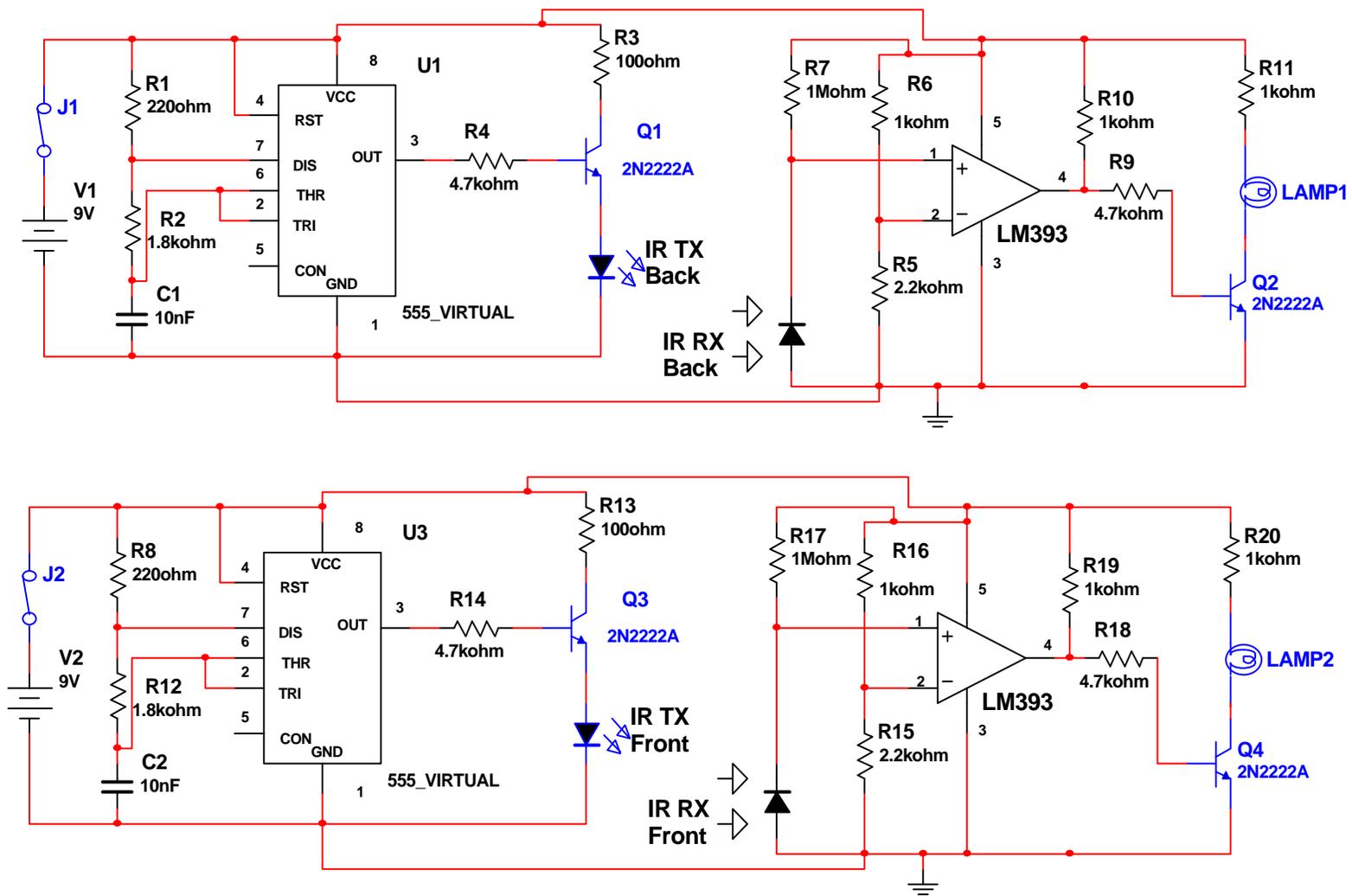

Fig. 13: Complete Circuit Diagram of Anti-collision System for Vehicles

# 4. SYSTEM TESTING, RESULTS AND DISCUSSIONS

Tests were carried out on the prototype device to see its effectiveness as to what degree it meets its expected performance. The construction of the system is in 2 stages, the soldering of the components and the coupling of the entire system to the casing. The outlet was wires intended to be connected to sensors mounted on each car, though a prototype of it was used instead. The power supply stage was first soldered, and then the transmitter and receiver stage and all the other stages were soldered. The circuit was soldered in a number of patterns that is, stage by stage. Each stage was tested using the multi-meter to make sure it is working properly before the next stage is done. This helps to detect mistakes and faults easily. The soldering of the circuit was done on a 10cm by 24cm Vero-board. The second stage of the system construction is the casing of the soldered circuit. This system was cased in a transparent plastic glass, this makes the system look attractive, and it helps in marketing the system because the circuit has to be attractive before someone would want to know what it does. The casing has special perforation and vent to ensure the system is not overheating, and this will aid in the life span of the circuit. Fig. 14 shows a prototype of the complete system.

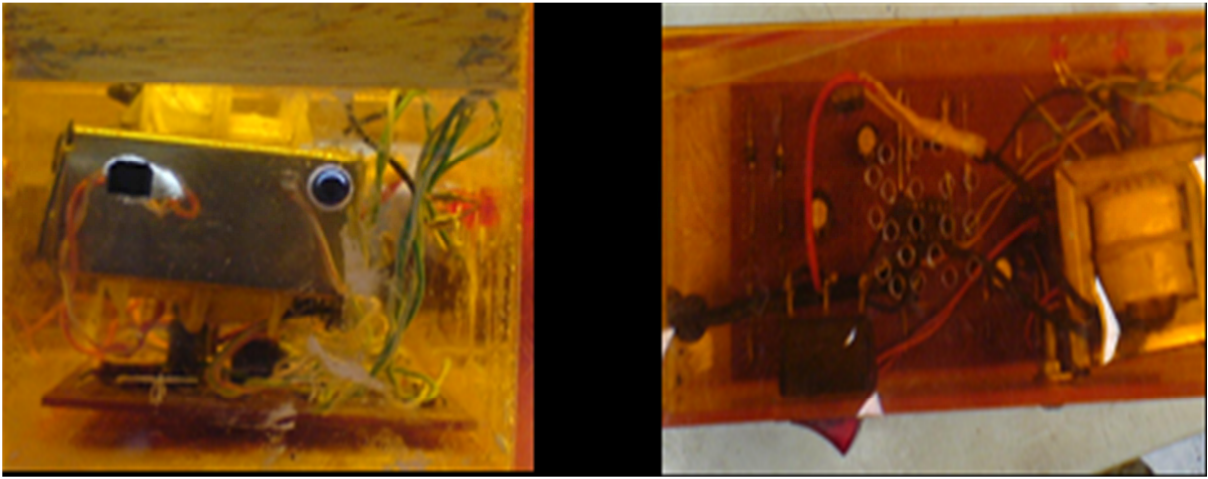

Fig. 14: Prototype of anti-collision model for vehicles

## 4.1. Testing

At first both devices are made to face each other to see if they would respond to the closeness. Then the distance between them was further increased to see the minimum within which it would still be effective as an anti–collision device. The physical realization of the system is very vital. This is where the fantasy of the whole idea meets reality. The designer will see his or her work not just on paper but also as a finished hardware. After carrying out all the paper design and analysis, the system was implemented and tested to ensure its working ability, and was finally constructed to meet desired specifications. The process of testing and implementation involved the use of some equipment such as digital Multimeter, signal generators and Oscilloscope and finally the physical testing for the device facing themselves to the transmitting and receiving ability.

## 4.2. Results

As they were brought close together it was noted that both devices had both LEDs 'ON' and their respective audio alerts came OFF. When they were brought to 0.4m of each other, the alarm ceased indicating that the circuits were working desirably. Although only one stopped sounding as the other had to be manually shut down to keep it from draining the battery.

### 4.3. Discussions

During the tests, it was noted that one of the devices kept signaling an alarm even when it had gone beyond the safe range (0.4m). This was due to the nature of the transmitter bought. There was a problem with purchasing a good one, as it was hard to tell from the point of purchase, which one would work and which would not, though, it was later replaced by another one and the whole system work perfectly. There was also a power supply problem as the speaker actually drained a lot of power from the battery, causing instability in the system thereby. Much of the testing was done with the speaker disconnected, and finally it was rectified as there was a short circuit on the board.

### 5. CONCLUSIONS

The system which is the design and construction of an anti-collision system for vehicles was designed considering some factors such as economy, availability of components and research materials, efficiency, compatibility, portability and also durability. The performance of the system after test met design specifications. The general operation of the system and performance is dependent on the presence of two moving cars as they get closer to each other. However, it should be stated here that the system was aimed at fabricating prototype, a replica of the actual thing. It is economically viable to undertake certain system this way since testing would not cost so much. Any desire to implement this design into a vehicle would require a laser detector. The problem of power supply would not arise due to the amount of battery power from the car battery. Also the operation of the system is dependent on how well the soldering is done, and the positioning of the components on the Vero board. The IC's were soldered away from the power supply stage to prevent heat radiation which, might occur and affect the performance of the entire system. The construction was done in such a way that it makes maintenance and repairs an easy task and affordable for the user should there be any system breakdown. All components were soldered on one Vero-board which makes troubleshooting easier. In general, the system was designed, and the real time implementation done with a photo-type of the model.

### ACKNOWLEDGEMENTS

The author would like to thank Col. Muhammed Sani Bello (RTD), OON, Vice Chairman of MTN Nigeria Communications Limited for supporting the research.

**Author**

**Engr. Adamu Murtala Zungeru** received his B.Eng. degree in Electrical and Computer Engineering from the Federal University of Technology (FUT) Minna, Nigeria in 2004, and M.Sc. degree in Electronic and Telecommunication Engineering from the Ahmadu Bello University (ABU) Zaria, Nigeria in 2009. He is a Lecturer Two (LII) at the Federal University of Technology Minna, Nigeria in 2005-till date. He is a registered Engineer with the Council for the Regulation of Engineering in Nigeria (COREN), Member of the Institute of Electrical and Electronics Engineers (IEEE), and a professional Member of the Association for Computing Machinery (ACM). He is currently a PhD candidate in the department of Electrical and Electronic Engineering at the University of Nottingham. His research interests are in the fields of automation, swarm intelligence, routing, wireless sensor networks, energy harvesting, and energy management for micro-electronics.

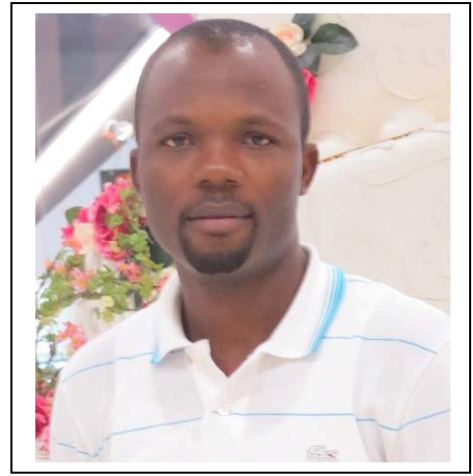